# Title

Afferent Fiber Activity-Induced Cytoplasmic Calcium Signaling in Parvalbumin-Positive Inhibitory Interneurons of the Spinal Cord Dorsal Horn

# Authors

Anna M. Hagenston*, Sara Ben Ayed, Hilmar Bading

Department of Neurobiology

Interdisciplinary Center for Neurosciences (IZN)

Heidelberg University

INF 366

69120 Heidelberg

Germany

*Corresponding author Anna M. Hagenston: hertle@nbio.uni-heidelberg.de

Sara Ben Ayed: khouaja@nbio.uni-heidelberg.de

Hilmar Bading: bading@uni-hd.de

## Abstract

Neuronal calcium ($Ca^{2+}$) signaling represents a molecular trigger for diverse central nervous system adaptations and maladaptions. The altered function of dorsal spinal inhibitory interneurons is strongly implicated in the mechanisms underlying central sensitization in chronic pain. Surprisingly little is known, however, about the characteristics and consequences of $Ca^{2+}$ signaling in these cells, including whether and how they are changed following a peripheral insult or injury and how such alterations might influence maladaptive pain plasticity. As a first step towards clarifying the precise role of $Ca^{2+}$ signaling in dorsal spinal inhibitory neurons for central sensitization, we established methods for characterizing $Ca^{2+}$ signals in genetically defined populations of these cells. In particular, we employed recombinant adeno-associated viral vectors to deliver subcellularly targeted, genetically encoded $Ca^{2+}$ indicators into parvalbumin-positive spinal inhibitory neurons. Using wide-field microscopy, we observed both spontaneous and afferent fiber activity triggered $Ca^{2+}$ signals in these cells. We propose that these methods may be adapted in future studies for the precise characterization and manipulation of $Ca^{2+}$ signaling in diverse spinal inhibitory neuron subtypes, thereby enabling the clarification of its role in the mechanisms underlying pain chronicity and opening the door for possibly novel treatment directions.

## Keywords

inhibitory neuron, dorsal spinal cord, calcium, central sensitization, GCaMP3, parvalbumin

## Main text

Changes in intracellular calcium ($Ca^{2+}$) levels in neurons are intimately linked to mechanisms underlying adaptive plastic changes within the central nervous system. $Ca^{2+}$ rises influence neuronal properties by diverse means, including modulation of ion channel activity or localization, regulation of neurotransmitter release, activation of $Ca^{2+}$-dependent enzymes, and induction or downregulation of gene expression. Dysregulation of intracellular $Ca^{2+}$ levels and their stimulus-evoked changes are similarly implicated in several central nervous system pathologies; aberrant intraneuronal $Ca^{2+}$ signaling is involved in processes as broad-ranging as excitotoxic neuronal death following stroke (1), impaired synaptic plasticity and memory formation in aging (2), and synapse loss in Alzheimer's disease (3), but also peripheral and central sensitization in chronic pain (4). Spinal central sensitization, defined as the aberrant amplification of noxious signal transmission in the spinal cord, is characterized both by the enhanced excitation of projection neurons and by a loss of inhibitory tone. Numerous mechanisms have been proposed to explain altered spinal inhibitory tone in central sensitization, including reduced excitation or excitability of inhibitory interneurons, degeneration of inhibitory neurons and/or their processes, reduced numbers of GABA- and glycinergic synaptic terminals, and diminished inhibitory neurotransmitter release (5, 6). Nonetheless, the precise molecular triggers underlying these alterations have not been elucidated. In light of its importance for regulating diverse neuronal functions, dysregulated intracellular $Ca^{2+}$ signaling clearly represents a possible proximal trigger for altered inhibitory neuron function in central sensitization. Surprisingly little is known, however, about the characteristics and consequences of $Ca^{2+}$ signaling in dorsal spinal inhibitory neurons, including whether it is altered in chronic pain states. We therefore aimed to establish methods for measuring $Ca^{2+}$ signals in defined inhibitory neuron types and subcellular compartments. Here we present data demonstrating the use of these methods for characterizing cytoplasmic $Ca^{2+}$ responses of parvalbumin (PV)-positive inhibitory neurons in the dorsal spinal cord to stimulation of sensory fiber afferents.

We used the genetically encoded calcium indicator (GECI) GCaMP3, linked to a nuclear export signal (NES), to characterize cytoplasmic $Ca^{2+}$ signals. While the maximum response amplitude and kinetics of GCaMP3 are inferior to those of, for example, GCaMP6f (7), this particular GECI has the advantage that its relatively high baseline fluorescence allows for the easy identification of transduced cells. Thus, GCaMP3 is our preferred GECI for wide-field Ca$^{2+}$ imaging in acute tissue preparations. We used recombinant adeno-associated viral (rAAV)-mediated gene delivery to express cytoplasmically-localized GCaMP3 (GCaMP3.NES) in PV-positive neurons of the spinal cord dorsal horn of PV-Cre mice. Specifically, we generated rAAV vectors containing a doubly floxed inverted open reading frame for GCaMP3.NES, which is expressed in a Cre recombinase-dependent manner under control of the human EF1α promoter (Fig. 1A). Initial tests carried out in primary hippocampal cultures co-infected (or not) with a second rAAV driving the expression of myc-tagged Cre recombinase—detected immunocytochemically—showed no GCaMP3.NES expression in the absence of Cre recombinase (data not shown). We injected rAAV vectors into the spinal parenchyma at spinal level L3/L4 and, after an incubation period of 3-4 weeks to allow for stable transgene expression, prepared acute transverse slices from the L3-L5 spinal cord with attached dorsal roots for Ca$^{2+}$ imaging (Fig. 1B). Consistent with numerous studies demonstrating that PV-positive inhibitory neurons are localized in lamina III and make projections into lamina II of the rodent spinal cord dorsal horn (8, 9), we observed GCaMP3.NES fluorescence in multiple cell bodies and processes within laminae II-III of the dorsal spinal cord (Fig. 1C).

Aβ, Aδ, and C afferent fibers, which carry information related to innocuous touch, fast pain, and slow pain, respectively, were electrically stimulated with varying frequency and amplitude through a suction electrode attached to the dorsal root. The chosen stimulation frequencies (4, 20, and 100 Hz) were roughly tuned to the maximum following frequencies of each fiber class (10, 11), and a broad range of stimulation intensities (0.01-2.0 mA) was employed in order to progressively activate each fiber class (12). As we have previously reported for excitatory neurons in laminae I-II recorded under similar conditions (13), PV-positive inhibitory neurons produced graded Ca$^{2+}$ responses to afferent fiber stimulation of increasing frequency and intensity (Fig. 1D,E; n=3-9 slices from 3-7 mice). PV-positive inhibitory neurons in lamina III predominantly receive synaptic input from Aβ sensory afferents (8). In keeping with this feature, we observed no sharp rise in response amplitudes above Aδ (~0.1 mA) or C (~0.5 mA) stimulation intensities for 20 Hz and 4 Hz stimulation trains, respectively. In most recorded slices (n=13/16 slices from 9 mice), PV-positive cell bodies and processes exhibited spontaneous $Ca^{2+}$ rises in the absence of synaptic input from the periphery (Fig. 1D), suggestive of their involvement in a basally active and recurrent spinal network. We have not observed such spontaneous activity in dorsal spinal excitatory neurons (data not shown).

The data we present here demonstrate that PV-positive inhibitory neurons in the dorsal spinal cord exhibit both afferent activity-evoked and spontaneous cytoplasmic $Ca^{2+}$ rises. Following peripheral nerve injury, the number of synapses formed by these cells onto excitatory interneurons decreases (9). Consequently, innocuous tactile information impinging on laminae III and IV gains access to nociceptive circuits in laminae I/II, and mechanical allodynia arises (9). A possible molecular mechanism underlying inhibitory synapse loss is the impaired activity- and $Ca^{2+}$-dependent expression of genes involved in inhibitory synapse maintenance (14). Thus, an examination of afferent activity-evoked $Ca^{2+}$ transients following nerve injury—embedded in a study employing gain- and loss-of-function tools specifically in PV-positive dorsal spinal neurons—may provide important mechanistic insight into spinal disinhibition in neuropathic pain.

We limited ourselves here to the description of $Ca^{2+}$ signaling in one inhibitory neuron type and subcellular compartment. Given the expanding repertoire of available Cre mouse lines for defined populations of inhibitory neurons, the broad range of proven subcellular targeting signals for GECIs, and the growing number of genetically encoded calcium signaling inhibitors (e.g., CaMBP4, PV.NLS (13, 15)), we envision that our methods can be easily adapted for studies aimed at gaining a precise understanding of the characteristics and consequences of spinal inhibitory neuronal Ca$^{2+}$ in central sensitization and pain chronification.

## Extended Materials and Methods

### *Mice*

Mice expressing the Cre recombinase in parvalbumin (PV)-positive inhibitory neurons (PV-Cre mice) were bred from founders originally obtained from Jackson Laboratories (JAX Stock #008069 (16)) and kindly provided by R. Kuner (Heidelberg University). All experimental procedures were approved by the local animal care and use committee (Regierungspräsidium Karlsruhe, Referat 35, Karlsruhe, Germany), project G-272/14, and carried out in accordance with ARRIVE guidelines and the EU Directive 2010/63/EU for animal experiments. Mice had access to food and water *ad libitum* and were housed on a 12 h light-dark cycle.

### *Production and intra-spinal injection of rAAV vectors*

The coding sequence for GCaMP3.NES, targeted to the cytoplasm using a nuclear export signal (NES), was subcloned using standard procedures between the NheI and AscI restriction sites of the Cre recombinase-dependent AAV expression plasmid, pAAV-EF1a-double floxed-EYFP-WPRE-HGHpA, a gift from Karl Deisseroth (Addgene plasmid # 20296; http://n2t.net/addgene:20296; RRID:Addgene_20296). Recombinant serotype 1/2 adeno-associated viral vectors (rAAV) were produced and purified using previously described methods (17, 18) and injected into the lumbar spinal cord of adult PV-Cre mice. Briefly, mice were anesthetized with an intraperitoneal injection of sleep mix containing 0.5 mg/kg medetomidine, 5.0 mg/kg midazolam, and 0.5 mg/kg fentanyl. When the mice were no longer responsive to a foot pinch, an ~1 cm incision was made in the skin over the vertebral column, and the tissue between thoracic vertebrae T12 and T13—corresponding to spinal lumbar segments L3/L4—was dissected to reveal the dorsal spinal cord. The dura mater was opened to improve needle penetration, and a 35 G beveled needle mounted in a 10 µl Hamilton syringe (World Precision Instruments) was slowly advanced into the spinal parenchyma. 500 nl of a 2:1 mixture of rAAV vector suspended in PBS and 20 % mannitol containing ~$10^{11}$ virions was then injected at a rate of 100 nl/min using a microprocessor-controlled minipump (World Precision Instruments). The needle was left in place for 5 min after the injection was complete to allow the viral infusion to diffuse away from the injection site. The wound was subsequently covered with gelatin foam (Gelita Medical) and the skin closed with sutures. Animals received carprofen (5 mg/kg s.c.) prior to the first incision and buprenorphine (0.1 mg/kg s.c.) 1 h after subcutaneous administration of wake mix containing 0.75 mg/kg atipamezole, 0.5 mg/kg flumazenil and 1.2 mg/kg naloxone to minimize post-surgical pain. After the procedure, mice were returned their home cage, which was placed on a heating plate for 24 h, and provided with wet food pellets. As we have demonstrated in the past (13), this viral delivery method generally results in transgene expression spreading from L2 to the border of S1 of the mouse spinal cord with a maximum in L3-L5, the area of the spinal cord that innervates the hindpaw.

### $Ca^{2+}$ imaging and dorsal root stimulation

Three to four weeks following rAAV injection surgery, acute spinal cord slices with attached dorsal roots from spinal cord segments L3-L5 of rAAV-injected mice were prepared as described previously (19). Briefly, mice were anesthetized with sodium pentobarbital (320 mg/kg, i.p.) and perfused with freshly prepared, ice-cold NMDG slicing solution containing, in mM, 93 HCl, 2.5 KCl, 1.2 $NaH_2PO_4$, 3 sodium pyruvate, 20 HEPES, 30 $NaHCO_3$, 25 D-glucose, 5 L-ascorbic acid, 10 N-acetyl-L-cysteine, 2 thiourea, 10 $MgSO_4$, 0.5 $CaCl_2$, adjusted to a final osmolarity of 300-305 mOsm and pH of 7.35, and bubbled with 95 % $O_2$/5 % $CO_2$. Afterwards, the spinal cord was isolated, the dura mater carefully removed, and the ventral roots resected. The spinal cord was then mounted into a groove within a block of 1.5

% agarose in PBS and secured with cyanoacrylate adhesive to the stage of a Microm HM650V (Thermo Scientific) vibratome containing NMDG slicing solution bubbled with 95 % $O_2$/5 % $CO_2$ and cooled to 1-4 °C. Slices (~330 μm) were prepared using double-edge coated blades (Science Services, washed version, #7200-WA) and then placed into NMDG slicing solution bubbled with 95 % $O_2$/5 % $CO_2$ and warmed to 34 °C for 10 min. Slices were subsequently transferred to extracellular recording solution containing, in mM, 125 NaCl, 10 D-glucose, 25 $NaHCO_3$, 2.5 KCl, 1.25 $NaH_2PO_4$, 1.2 $MgSO_4$, 2 $CaCl_2$, adjusted to a final osmolarity of 300-305 mOsm and pH 7.35, bubbled with 95 % $O_2$/5 % $CO_2$ and warmed to 34 °C, and then allowed to cool to room temperature for a recovery period of ≥ 1 h. For $Ca^{2+}$ imaging, slices were transferred to a recording chamber (PM-1, Warner Instruments, Hamden, CT, USA or PC-R, Siskiyou, OR, USA) perfused (~3 ml/min) with 95 % $O_2$/5 % $CO_2$-bubbled extracellular recording solution warmed with an in-line heater to 34 °C. Slices were secured with a platinum ring with nylon strings. Dorsal roots were sucked into a glass stimulation electrode, the tip of which had been dipped into Vaseline to improve the electrical seal around the root, and stimulated with 1 s trains of electrical pulses (0.1 ms pulse width, 100 Hz and 20 Hz stimuli; 1 ms pulse width, 4 Hz stimuli) using a stimulus isolator (World Precision Instruments, A365) controlled by Clampex 10.3 Software (Molecular Devices). $Ca^{2+}$ responses from GCaMP3.NES-labeled cells and their processes in laminae II-III were imaged as described in detail previously (20). GCaMP3.NES was excited using a CoolLED light source (480±10 nm), and fluorescence emission (530±20 nm) acquired at a rate of 2 Hz with a cooled camera (Photometrics Coolsnap HQ, Roper Scientific) with 2x2 binning through a 20x water-immersion objective (XLM PlanFluor 0.95W, Olympus) on an upright microscope (Olympus BX51WI) connected to a software interface (Metafluor, Universal Imaging Systems and Molecular Devices). Overview images were taken with a 4x air objective (PlanCN, Olympus).

*Data analysis*

Image sequences were imported into ImageJ (Fiji RRID:SCR_002285) and, when necessary, processed with the "Template Matching" plugin to correct for slice movement during the recording. ROIs were manually drawn around single cells and larger regions encompassing nearly the entire field of view within laminae II/III containing fluorescently labeled cells and processes, and fluorescence intensity changes measured over time. Further data analysis was carried out using Igor Pro (WaveMetrics, RRID:SCR_000325). Fluorescence intensity changes were expressed as the percent change with respect to baseline in the 15 s prior to the onset of the stimulation train (% ΔF/F) and quantified using the peak amplitude above baseline. Data were plotted as mean ± SEM using GraphPad Prism 6 (RRID:SCR_002798).

*Post-imaging analysis of GCaMP3.NES expression*

Following $Ca^{2+}$ imaging, acute spinal cord slices were fixed in 4 % paraformaldehyde in PBS for 2 h than them placed in 30 % sucrose in PBS containing 0.04 % Thimerosal (Sigma-Aldrich) for a minimum of 24 h. Tissue slices were embedded in tissue freezing medium (Leica), and 20 μm sections cut using a cryotome (CM 1950, Ag Protect, Leica) and mounted directly onto SuperFrost Plus slides (Thermo Scientific). Slides were coverslipped with Mowiol 4-88 (Calbiochem) medium containing 2 μg/ml Hoechst 33258 (Serva, Cat # 15090) as a nuclear counterstain. Images of the dorsal horn were obtained with an upright wide-field microscope (DM IRBE, Leica) and 20x objective and processed using Adobe Photoshop CS4.

**List of abbreviations**

$Ca^{2+}$: calcium

GECI: genetically encoded calcium indicator

NES: nuclear export signal

PV: parvalbumin

rAAV: recombinant adeno-associated virus

## Declarations

### *Ethics approval*

This study was approved by the local animal care and use committee (Regierungspräsidium Karlsruhe, Referat 35, Karlsruhe, Germany), project G-272/14. All animal experiments were designed and carried out in accordance with ARRIVE guidelines and the EU Directive 2010/63/EU for animal experiments.

### *Availability of data and materials*

The data sets analyzed during the current study are available from the corresponding author upon reasonable request.

### *Consent for publication*

Not applicable

### *Competing interests*

The authors declare that they have no competing interests.

### *Funding*

This study was funded by the German Research Foundation (SFB1158, grant A05).

### *Acknowledgements*

We would like to thank Rohini Kuner (Heidelberg University) for providing PV-Cre mice for use in this study.

*Authors' contributions*

AMH and HB conceived the project, designed the study, and wrote the paper. AMH and SK performed the experiments. AMH and HB analyzed and interpreted the data. All authors read and approved the final manuscript.

## Figures

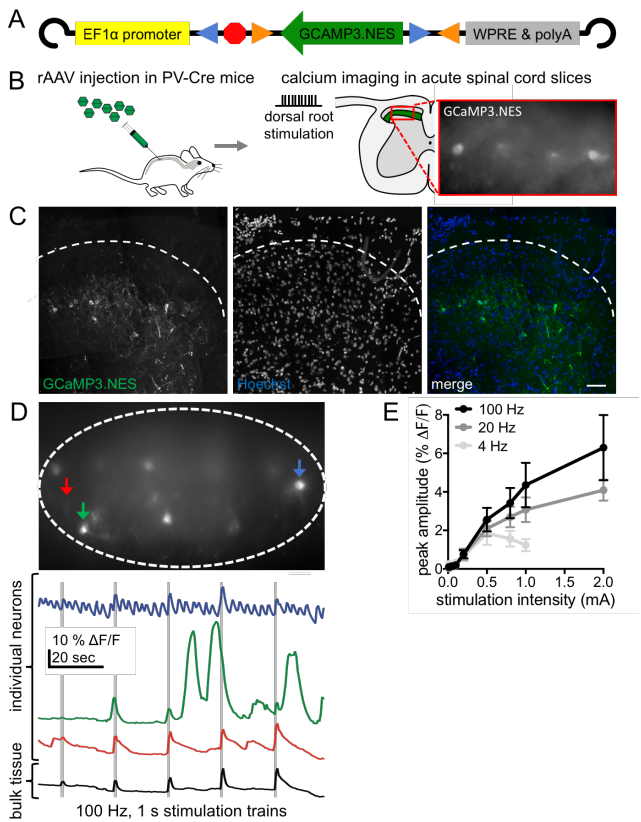

***Figure 1.*** Afferent activity-triggered and spontaneous Ca$^{2+}$ signaling in dorsal spinal inhibitory interneurons. (A) We generated serotype 1/2 rAAV vectors containing a doubly floxed inverted open reading frame coding for the cytoplasmically-localized GECI, GCaMP3.NES, expressed in a Cre recombinase-dependent manner under control of the EF1α promoter. (B) rAAV vectors were injected into the spinal parenchyma of adult PV-Cre mice. Acute slices with attached dorsal roots from the L3-L5 spinal cord were prepared 3-4 weeks later. Sensory afferent activity was evoked by suction electrode stimulation of the attached dorsal root. Recordings of Ca$^{2+}$ responses from PV-positive neurons were made in laminae II-III of the spinal cord dorsal horn. (C) Analysis of GCaMP3.NES expression in the spinal cord dorsal horn. GCaMP3.NES was observed in the extranuclear soma and processes of cells residing in lamina III and extending processes into lamina II. The upper border of lamina I is marked with a dashed line. Scale bar, 100 µm. (D) Top, representative image from a Ca$^{2+}$ imaging experiment in which multiple GCaMP3.NES-positive neurons

could be visualized. Bottom, changes in cytoplasmic Ca$^{2+}$ levels were visualized using the normalized fluorescence intensity in regions of interest laid over individual neurons (red, green, and blue traces; corresponding the colored arrows in the image above) and the bulk tissue (black traces; corresponding to the large white ellipse in the image above) during 100 Hz dorsal root stimulation (0.2-2.0 mA). Both stimulus-evoked responses and spontaneous activity with highly variable amplitude and kinetics were observed. (E) Quantification of Ca$^{2+}$ rises in the bulk tissue triggered by 1 s, 0.01-2.0 mA stimulation trains delivered at 4, 20, and 100 Hz (n = 3-9 slices from 3-7 animals). The amplitudes of evoked Ca$^{2+}$ transients varied positively with both stimulus frequency and intensity.